\PassOptionsToPackage{table,xcdraw}{xcolor}

\documentclass[10pt,conference]{IEEEtran}
\usepackage{alltt                                    % I like these
	, multirow
	,subfig
	, booktabs
	, listings
	, graphicx
	,float
	,cite
	,verbatim
	,mathtools
	,url
}

\usepackage{xcolor}
\usepackage{xcolor, soul}
\usepackage[numbers]{natbib}
\usepackage{threeparttable}
\usepackage{framed}
\usepackage{expl3}
\usepackage{hyperref}
\usepackage{filecontents}

\usepackage{pgfplots}
\pgfplotsset{compat=1.16}
\usepackage{pgfplotstable}

\usepgfplotslibrary{statistics}

\usepackage{scalefnt}
\usepackage{tikz}

\usepackage{tabularx}
\usepackage{lipsum}
\usepackage{array}

\usepackage[most]{tcolorbox}
\usepackage{multirow}

\usetikzlibrary{shapes.gates.logic.US,trees,positioning,arrows}
\usetikzlibrary{shapes.multipart}
\usetikzlibrary{shapes,arrows}

\tikzstyle{vertex}=[ellipse,fill=black!25,minimum size=20pt, inner sep=0pt]
\tikzstyle{edge} = [draw,thin,-]
\tikzstyle{glabel} = [text width=1cm,text centered,font=\bf]
\pgfdeclarelayer{bg}    % declare background layer
\pgfsetlayers{bg,main} 

\ExplSyntaxOn
\newcommand\latinabbrev[1]{
  \peek_meaning:NTF . {% Same as \@ifnextchar
    #1\@}%
  { \peek_catcode:NTF a {% Check whether next char has same catcode as \'a, i.e., is a letter
      #1., \@ }%
    {#1., \@}}}
\ExplSyntaxOff

\usepackage{tikz}
\usepackage{verbatim}
\usetikzlibrary{arrows,shapes,backgrounds}

\tikzstyle{vertex}=[ellipse,fill=black!25,minimum size=20pt, inner sep=0pt]
\tikzstyle{edge} = [draw,thin,-]
\tikzstyle{glabel} = [text width=1cm,text centered,font=\bf]
\pgfdeclarelayer{bg}    % declare background layer
\pgfsetlayers{bg,main}  

\usepackage{soul}
\usepackage{pifont}
\usepackage{booktabs,makecell}
\usepackage[export]{adjustbox}
\usepackage{balance}
\usepackage{paralist}

\usepackage{pgf-pie}

\newif\ifpienumberinlegend
\pgfkeys{/number in legend/.code=
    \expandafter\let\expandafter\ifpienumberinlegend
    \csname if#1\endcsname
    \ifpienumberinlegend

    \def\beforenumber##1\afternumber{}%
    \fi,
    /number in legend/.default=true
}

\makeatletter
\pgfplotsset{
    boxplot/hide outliers/.code={
        \def\pgfplotsplothandlerboxplot@outlier{}%
    }
}
\makeatother

\newcommand{\CASE}[1]{\STATE \textbf{case} #1\textbf{:} \begin{ALC@g}}
\newcommand{\ENDCASE}{\end{ALC@g}}

\newcommand{\DEFAULT}{\STATE \textbf{default:} \begin{ALC@g}}
\newcommand{\ENDDEFAULT}{\end{ALC@g}}
\newcommand{\DEFAULTLINE}[1]{\STATE \textbf{default:} }
\let\footnotesize\scriptsize

\newsavebox{\supbox}% Superscript box
\newcommand{\bsup}{\begin{lrbox}{\supbox}$\tt\scriptstyle}% Superscript begin
\newcommand{\esup}{$\end{lrbox}{}^{\usebox{\supbox}}}% Superscript end

\definecolor{lightpurple}{rgb}{0.8,0.8,1}
\definecolor{codebg}{RGB}{255,255,255}
\definecolor{commentcolor}{RGB}{11,140,11}
\usepackage{graphicx}  % Required for rotation
\tcbset{valign=center, add to natural height=0.4cm, blanker, borderline west={1.5mm}{0mm}{gray},  interior engine=spartan, colback=gray!10, add to height=1cm}
\newtcolorbox[]{findingbox}[1][] { #1 }
\newtcolorbox[]{promptbox}[1][] { reset, #1}

\definecolor{mygrayzero}{gray}{0.9}
\definecolor{mygrayone}{gray}{0.8}
\definecolor{mygraytwo}{gray}{0.7}
\definecolor{mygraythree}{gray}{0.6}

\def\BibTeX{{\rm B\kern-.05em{\sc i\kern-.025em b}\kern-.08em
    T\kern-.1667em\lower.7ex\hbox{E}\kern-.125emX}}

\usepackage[normalem]{ulem} % for \sout
\usepackage{xcolor}

\usepackage{ifthen}
\usepackage{amssymb}
\newboolean{showcomments}

\setboolean{showcomments}{false}
\ifthenelse{\boolean{showcomments}}
{\newcommand{\nbc}[3]{
 {\colorbox{#3}{\bfseries\sffamily\scriptsize\textcolor{white}{#1}}}
 {\textcolor{#3}{\sf\small$\blacktriangleright$\textit{#2}$\blacktriangleleft$}}
 }
}
{\newcommand{\nbc}[3]{}
 }

\usepackage{amsmath}
\usepackage{enumitem}

\begin{document}

% \title{Predicting Unanswered Questions of SO During their Submission}
% \title{Early Detection of Unanswered Questions and Evidence-based Guideline to Increase Their Answerability}

% \title{Does Your Question at SO Need a Code Snippet? Investigating the Cause \& Effect of Missing Code Snippets}
% \title{Impact of Editing on Code Quality and LLMs' Role in Automated Quality Estimation on Technical Q\&A Sites}
% \title{Does Editing Improve Stack Overflow Answer Quality? An Empirical Study on Editing's Effect and LLM's Ability to Estimate Code Quality}
\title{Does Editing Improve Answer Quality on Stack Overflow? A Data-Driven Investigation}

\author{Anonymous Authors}
\author{
\IEEEauthorblockN{Saikat Mondal}
\IEEEauthorblockA{University of Saskatchewan, Canada\\
\ saikat.mondal@usask.ca}
\and
% \IEEEauthorblockN{Mohammad Masudur Rahman}
% \IEEEauthorblockA{Dalhousie University\\
% \ masud.rahman@dal.ca}
% \and
\IEEEauthorblockN{Chanchal K. Roy}
\IEEEauthorblockA{University of Saskatchewan, Canada\\
\ chanchal.roy@usask.ca}
}

% \author{Ben Trovato}
% \authornote{Both authors contributed equally to this research.}
% \email{trovato@corporation.com}
% \orcid{1234-5678-9012}
% \author{G.K.M. Tobin}
% \authornotemark[1]
% \email{webmaster@marysville-ohio.com}
% \affiliation{%
%   \institution{Institute for Clarity in Documentation}
%   \streetaddress{P.O. Box 1212}
%   \city{Dublin}
%   \state{Ohio}
%   \postcode{43017-6221}
% }

%PredUnations

% \author{\IEEEauthorblockN {Saikat Mondal\hspace{0.7cm}Avijit Bhattacharjee\hspace{0.7cm}Chanchal K. Roy}
% \IEEEauthorblockA{SRlab, Department of Computer Science, University of Saskatchewan\\ }}
% \ saikat.mondal@usask.ca}
% }

% \renewcommand{\shortauthors}{Saikat Mondal et al.}
% \renewcommand{\shorttitle}{Reproducibility Challenges and Their Impacts on Technical Q\&A Websites}

\maketitle

\begin{abstract}

High-quality answers in technical Q\&A platforms like Stack Overflow (SO) are crucial as they directly influence software development practices. Poor-quality answers can introduce inefficiencies, bugs, and security vulnerabilities, and thus increase maintenance costs and technical debt in production software. To improve content quality, SO allows collaborative editing, where users revise answers to enhance clarity, correctness, and formatting. Several studies have examined rejected edits and identified the causes of rejection. However, prior research has not systematically assessed whether accepted edits enhance key quality dimensions. While one study investigated the impact of edits on C/C++ vulnerabilities, broader quality aspects remain unexplored.
In this study, we analyze 94,994 Python-related answers that have at least one accepted edit to determine whether edits improve (1) semantic relevance, (2) code usability, (3) code complexity, (4) security vulnerabilities, (5) code optimization, and (6) readability. Our findings show both positive and negative effects of edits. While 53.3\% of edits improve how well answers match questions, 38.1\% make them less relevant. Some previously broken code (9\%) becomes executable, yet working code (14.7\%) turns non-parsable after edits. Many edits increase complexity (32.3\%), making code harder to maintain. Instead of fixing security issues, 20.5\% of edits introduce additional issues. Even though 51.0\% of edits optimize performance, execution time still increases overall. Readability also suffers, as 49.7\% of edits make code harder to read.
This study highlights the inconsistencies in editing outcomes and provides insights into how edits impact software maintainability, security, and efficiency that might caution users and moderators and help future improvements in collaborative editing systems.

\end{abstract}


%________________________
\section{Introduction}
\label{sec:introduction}
%________________________
The quality of answers on Technical Q\&A sites like Stack Overflow (SO) is crucial since it directly impacts software development. However, the initial versions of many answers often contain inaccuracies, inefficiencies, or bugs, and thus make them suboptimal. These low-quality answers hurt the development of high-quality, optimized, and usable software and increase technical debt and maintenance costs. Therefore, SO allows users to improve these answers by suggesting edits. After an answer is submitted, it can be self-edited by the answerer and/or collaboratively edited by some editors (i.e., users other than the answerer). As of February 18, 2025, 11.2 million answers (i.e., 31.1\% of all about 36 million answers) have been edited at least once. In SO, edits by trusted contributors (i.e., users with 2000+ reputation scores) and post owners will be directly accepted. Edits by other contributors will be accepted only if at least two of the three trusted contributors approve them. 
While users sometimes make complex edits (e.g., adding or removing sentences and code snippets), they predominantly perform simpler edits focused on correcting misspellings, fixing grammar errors, and formatting to align with conventions or SO guidelines \cite{li2015good, mamykina2011design, wang2018users}. 
However, the critical question remains: \textit{Do these edits genuinely enhance objective quality?} It is vital to investigate whether users prioritize minor edits over real improvements, as this might compromise software quality.

\begin{figure}[!htb]
	\centering
	\includegraphics[width = 3.4in]{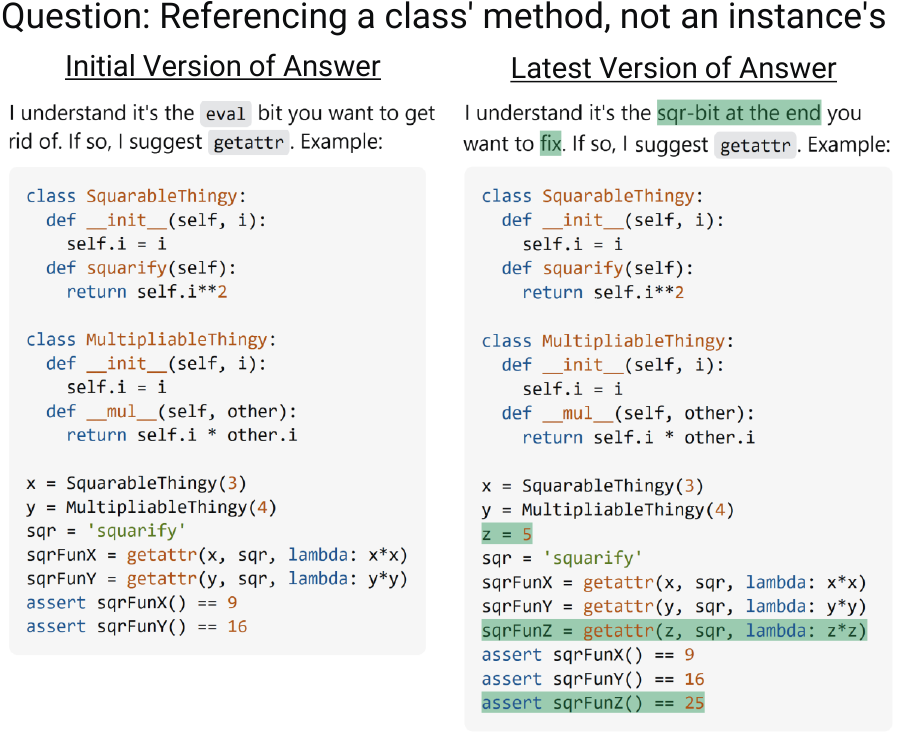}
	\caption{An example edits of Stack Overflow answer.}
	\label{fig:example}
	% \vspace{-5mm}
\end{figure}

Consider the example answer edits\footnote{\url{https://stackoverflow.com/posts/950139/revisions}} in Fig. \ref{fig:example}, where the editor makes minor text edits and adds three lines of code. However, the initial code had two potential instances of \textit{CWE-703: Improper Check or Handling of Exceptional Conditions} due to the unsafe use of \texttt{getattr()} with a default \textit{lambda} function that assumes multiplication will always work. The latest version introduces another \textit{CWE-703} by applying the same risky pattern to an integer $(z)$, which could cause unexpected errors if $z$ were not an integer. Without resolving the previous two CWE-703 issues, the edits introduce an additional risk, further increasing the likelihood of failure.
This example highlights a key issue -- edits may perform minor changes but can overlook quality threats and even introduce new ones. It aligns with our study's goal of comprehensively assessing whether edits truly enhance answer quality or introduce unintended risks in security, complexity, and maintainability.

Through a qualitative study, Wang et al. \cite{wang2018users} identified 12 reasons (e.g., undesired code/text formatting) behind rejected edits by rollbacks, while Mondal et al. \cite{mondal2023automatic} added seven more (e.g., adding gratitude). Although these studies provide insights into rejection reasons, they do not assess whether accepted edits improve critical quality aspects of answers, such as readability, code complexity, and code vulnerability—factors essential for efficient, reliable, and maintainable software development. Zhang et al. \cite{zhang2021study} explore whether edits reduce Common Weakness Enumeration (CWE) vulnerabilities in C/C++ code snippets. Their findings suggest that while edits slightly mitigate code weaknesses, most issues remain unaddressed. These findings highlight the need for a comprehensive study that evaluates all major aspects of answer quality, which can provide deeper insights to SO users who share and edit answers and help them recognize potential quality issues.

In this study, we conduct an empirical study to analyze whether edits address the critical quality aspects of SO answers.
We first extracted 95,556 answers related to the Python programming language with at least one accepted revision. Next, we evaluate their quality by applying 12 objective metrics across six key aspects: (a) semantic relevance, (b) code usability, (c) code complexity, (d) code vulnerabilities, (e) code optimization, and (f) readability. We compare these metric values between the initial and latest versions of the answers to see quality improvements.
% We then select XX metrics (out of 11) explicitly related to code quality and estimate their values using GPT-4o. To achieve this, we randomly sampled XX examples from the latest version of answers with executable code snippets and fine-tuned GPT-4o with XX labeled samples. We evaluated its performance using YY test samples. This experiment aims to assess to what extent LLMs can estimate code quality accurately, and its success could unlock new possibilities for future LLM-powered tools to consolidate multiple evaluation metrics into a single, efficient solution.
In particular, we answer six research questions through six major contributions in this paper, as follows:

\begin{description}
    \item[RQ1 (\textit{Semantic Relevance}):] \textbf{Do edits enhance the semantic similarity between questions and answers, ensuring greater relevance and coherence?}
    Semantic relevance is essential for maintaining strong alignment between questions and answers to enable clear and accurate knowledge transfer. Edits that improve semantic similarity can boost relevance and coherence and enhance the overall quality of the answer. We thus assess whether edits effectively strengthen this alignment to support more effective information exchange.
    % Semantic Relevance Metric
    % Question-Answer Similarity: Measures how relevant and aligned an answer is with its corresponding question.

    \item[RQ2 (\textit{Code Usability}):] \textbf{Do edits improve code usability by enhancing its parsability and executability?}
    Code usability (i.e., parsability and executability) is crucial because it directly impacts a developer's ability to integrate, test, and deploy shared code snippets efficiently. We thus analyze whether edits enhance the parsability and executability of answer code snippets.
    % Code Usability Metrics
    % Parsability of Code: Ensures that code snippets are syntactically correct and can be parsed without errors.
    % Compilability/Executability of Code: Assesses whether the code compiles and runs successfully.?   

    \item[RQ3 (\textit{Code Complexity}):] \textbf{Do edits contribute to reducing code complexity, making it easier to maintain and modify?}
    Code complexity, measured through metrics such as cyclomatic complexity, Halstead complexity, and the maintainability index, significantly influences a developer's ability to understand, maintain, and adapt code efficiently. Therefore, we assess whether edits effectively reduce complexity to promote enhanced maintainability and modifiability of answer code snippets.

    % Code Complexity Metrics
    % Cyclomatic Complexity: Measures the number of independent paths through the code.
    % Halstead Complexity: Evaluates the complexity of code based on operators and operands.
    % Maintainability Index: Determines how easy the code is to maintain and modify.

    \item[RQ4 (\textit{Code Vulnerabilities}):] \textbf{Do edits reduce security vulnerabilities, bugs, and code smells in Stack Overflow answers?}
    Code quality is not only determined by functionality but also by its robustness against security vulnerabilities, bugs, and code smells. Such issues can lead to security risks, maintenance challenges, and unreliable software behavior. Therefore, we examine whether edits effectively mitigate these risks by enhancing the security and quality of answer code snippets.
    % Security, Bug,  and Code Smell Metrics
    % Security using Bandit: Detects potential security vulnerabilities in Python code.
    % Bugs and Code Smells using SonarQube: Identifies potential defects and code quality issues, such as duplicated code, long methods, and improper exception handling. 

    \item[RQ5 (\textit{Code Optimization}):] \textbf{Do code edits in Stack Overflow answers make them more optimized by reducing the number of function calls and time complexity?}
    Code optimization is not solely defined by correctness, but it also involves improving execution efficiency. Inefficient code can cause performance bottlenecks and scalability challenges. Therefore, we examine whether edits in Stack Overflow answers enhance efficiency by reducing function calls and execution time.

    \item[RQ6 (\textit{Readability}):] \textbf{Do edits improve the readability of Stack Overflow answers to make them easier to comprehend for users?}
    Readability ensures that texts and codes of SO answers are easily understood and accessible to users. High readability can enhance learning, reduce confusion, and support quicker problem-solving. We thus investigate whether edits contribute to improving readability to help users grasp solutions more effectively.
    % Readability  Metrics
    % Text Readability: Evaluates how easily the textual content of an answer can be read and understood.
    % Code Readability: Evaluates the clarity of the code in terms of formatting, variable naming, indentation, and structure.

    % \item[RQ7 (\textit{AI-Driven Code Quality Judgment}):] \textbf{Can LLMs (e.g., GPT) reliably estimate the objective quality of code snippets in Stack Overflow answers?}
    % Large Language Models like GPT demonstrated potential in code generation and completion tasks. Reliable AI-driven code quality assessments could enhance automated code review, recommendation systems, and quality assurance. We thus explore whether LLMs can estimate code quality effectively to enable the development of future tools that align with established code quality standards.
    
\end{description}

\noindent The findings of our study can make SO users and moderators aware and help forum designers improve collaborative editing systems and guidelines to ensure that edits improve crucial answer quality (e.g., usability of the code, complexity, vulnerabilities, and readability), leading to more reliable knowledge sharing.

\smallskip
\noindent\textbf{Replication Package} can be found in our online appendix \cite{replicationPackage}.

\begin{figure*}[t]
	\centering
	\includegraphics[width = 5.5in]{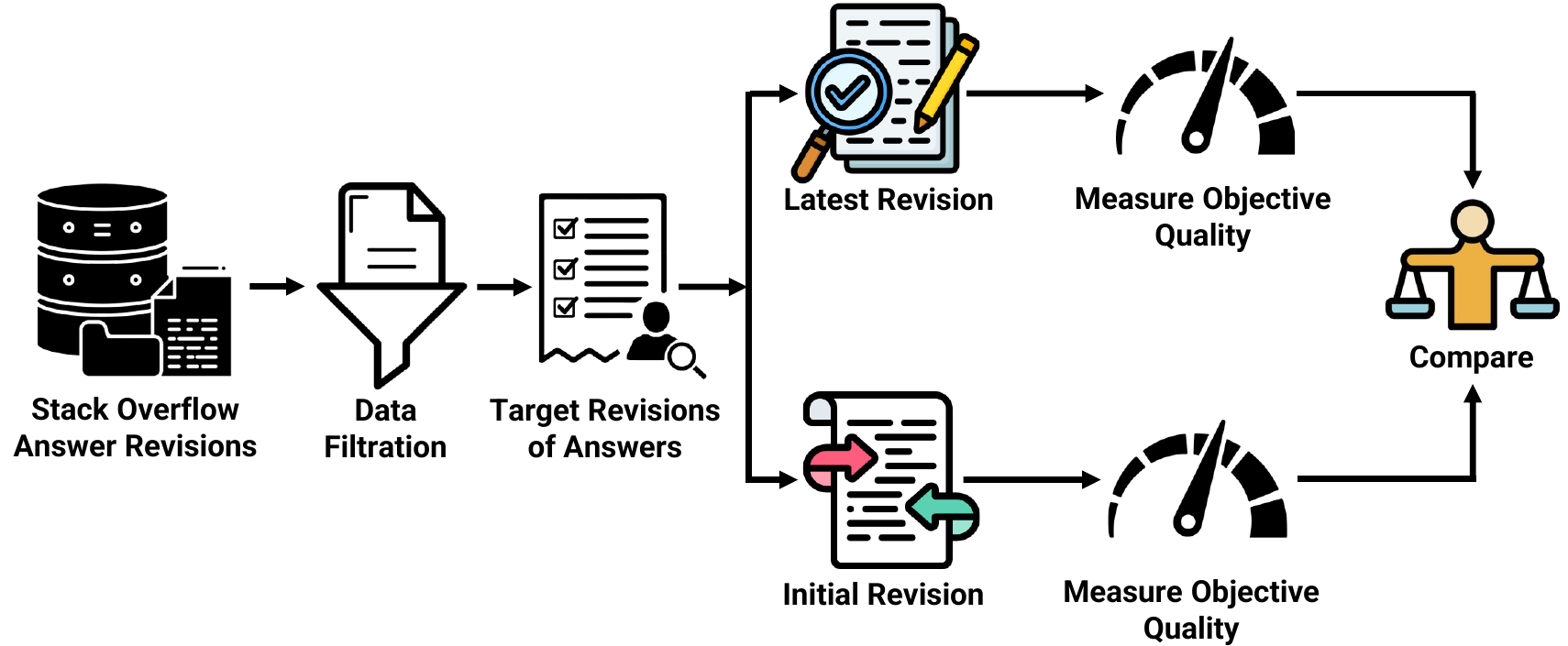}
	\caption{Study methodology}
	\label{fig:studyMethodology}
	% \vspace{-4mm}
\end{figure*}

% \vspace{-2mm}

\section{Study Methodology}
\label{sec:methodology}

Fig. \ref{fig:studyMethodology} shows the schematic diagram of our study methodology.  
The following sections discuss the different steps of our methodology.

\subsection{Dataset Construction}
\label{subsec-mothodology:createDataset}

We collected the January 2025 data dump of SO from the Stack Exchange site \cite{datadumpapi}. We aim to investigate the impact of editing on several quality attributes of SO answers. Our focus is on answers related to Python, which is one of the most widely used programming languages. We specifically chose Python for this study due to its suitability for evaluating objective metrics that require compilable/executable code. While static programming languages like Java have only 1.0\% of answers with executable code \cite{querytousablecode}, Python’s dynamic typing and flexibility allow us to capture more executable code.
This makes Python a better choice for large-scale analysis and helps us effectively assess the impact of editing on answer quality using a broad set of quality perspectives.

We selected answers to questions tagged with Python or related tags (e.g., Python-2.7, Python-3.6). From these, we identified 95,556 answers that met our selection criteria (i.e., had at least one accepted edit).
To analyze whether users consider quality when editing answers, we collected two versions of answers: the initially submitted version and the latest version after revisions. The latest versions are available in the SO data dump. To obtain the initial versions, we saved web pages containing the revision history, parsed the HTML, and extracted the initial versions of each answer. We then compute the edit distance between the initial and latest versions of answers using the Levenshtein distance \cite{levenshtein1965binary}. Our analysis finds that 0.6\% (562 out of 95,556) of the answer pairs have an edit distance of zero, which indicates no visible changes in the latest revisions. This scenario occurs when intermediate revisions introduce changes that are subsequently reverted, leaving the latest version unchanged from the initial one. We excluded these cases from our analysis. This resulted in a final dataset of 94,994 answers with modifications to either the text or code snippets (if present).

\subsection{Semantic Relevance Assessment (RQ1)}
\label{subsec-mothodology:relevance}

We evaluate whether edits improve the relevance of answers to questions by comparing the initial and latest versions of answers against the corresponding questions. We apply Sentence-BERT (SBERT) \cite{reimers2019sentence}, a refined version of the pre-trained BERT model that employs siamese and triplet network structures to generate semantically meaningful sentence embeddings. Unlike traditional metrics such as ROUGE and BLEU, which rely on surface-level n-gram overlaps, SBERT captures deep contextual and semantic relationships between texts and has demonstrated proven effectiveness in semantic textual similarity (STS) tasks \cite{ha2021utilizing}.

For this study, we use the \textit{all-mpnet-base-v2}\footnote{https://huggingface.co/sentence-transformers/all-mpnet-base-v2} model, which maps sentences into a 768-dimensional dense vector space. Our analysis includes 94,994 samples of the initial and latest versions of answers (see Section \ref{subsec-mothodology:createDataset}) and their corresponding questions. 
We consider the entire content of both answers and questions (title + body), including code (if any), to measure similarity. 
We calculate cosine similarity scores (0--1) between (a) the initial answer and the question and (b) the latest answer and the question. A smaller angle between vectors indicates higher similarity, while a larger angle suggests lower similarity. Finally, we compute the difference between the two similarity scores to quantify the impact of edits on answer relevance.

\subsection{Code Usability Evaluation (RQ2)}
\label{subsec-mothodology:usability}

We analyze code parsability to determine whether edits improve the usability of initially submitted code snippets. In this study, we define usability based on the standard processes of parsing and executing source code. We extract code snippets from answers using specialized HTML tags like $<$code$>$ within $<$pre$>$.
Our analysis finds that 79,415 initial answer versions contain code snippets, while the number is 85,993 for the latest versions. We found that edits included code snippets to 6,789 answers that initially lacked them. On the contrary, 211 answers had code snippets in their initial versions, which were later removed through edits. However, 79,204 answers contain code snippets in both the initial and latest versions. We then compute the edit distance between the initial and latest code snippets using the Levenshtein distance. The results show that 75.2\% (59,561 out of 79,204) of code snippets have an edit distance greater than zero, indicating actual modifications. The remaining snippets remain unchanged. We thus focus on the 59,561 answers with edited code snippets to assess whether edits enhance code usability.

We use Python's built-in AST module (e.g., ast.parse(code)) to parse code strings. We evaluate the executability of code snippets within a controlled environment with a 60-second time limit to prevent unresponsive processes. Code snippets are stored as temporary files and executed using Python's subprocess module.
During initial experiments, the execution was interrupted by potentially harmful code containing shutdown commands, process termination, or destructive file deletion operations. 
To address this, we implemented heuristic-based filtering to detect and skip such risky code to ensure a seamless and controlled experiment. 
In particular, we filtered out code snippets containing risky patterns such as infinite loops (e.g., \texttt{while 1}, \texttt{while True}), dangerous system commands (e.g., \texttt{rm -rf}, \texttt{system shutdown}), and process termination calls (e.g., \texttt{kill(PID)}, \texttt{killpg(PGID)}, \texttt{subprocess.kill}, or \texttt{import os} combined with \texttt{kill}). 
However, these unsafe code snippets represent less than 1\% of our dataset and thus do not affect the validity of our findings.

% We merge code snippets when answers contain multiple code blocks. Since Python is an interpreted language, this merging does not affect code parsability or executability and the validity of our findings.

\subsection{Code Complexity Measurement (RQ3)}
\label{subsec-mothodology:complexity}

Software complexity is a critical factor that degrades quality by increasing the difficulty of comprehension and maintenance \cite{malhotra2015python}. This study evaluates three core complexity metrics: Cyclomatic Complexity, Halstead Complexity, and the Maintainability Index.
\textit{Cyclomatic complexity} \cite{weyuker1988evaluating} (established by Thomas J. McCabe \cite{mccabe1976complexity}) quantifies the number of linearly independent paths in a program and reflects its decision logic \cite{malhotra2015python}.
\textit{Halstead Complexity} is an important technique to measure code complexity based on the frequency of operands and operators \cite{basili1984software, naveed2021measuring, alfadel2017evaluation}.
In particular, we compute \textit{volume}, \textit{difficulty}, and \textit{effort} as follows:

\begin{itemize}
    \item \textbf{Volume (V)} measures the size of the implementation, representing the number of bits required to encode the program:
    % \begin{equation}
        $V = N \log_2 \eta$
    % \end{equation}

    \item \textbf{Difficulty (D)} represents how challenging the code is to understand and modify:
    % \begin{equation}
        $D = \frac{\eta_1}{2} \times \frac{N_2}{\eta_2}$
    % \end{equation}

    \item \textbf{Effort (E)} estimates the mental effort required to write and understand the code:
    % \begin{equation}
        $E = D \times V$
    % \end{equation}
\end{itemize}

\noindent where $\eta_1$ = Number of distinct operators,
    $\eta_2$ = Number of distinct operands,
    $N_1$ = Total number of operators, and
    $N_2$ = Total number of operands.
The \textit{Maintainability Index} integrates multiple traditional source code metrics into a single value that indicates the relative ease of code maintenance and modification \cite{welker2001software, wani1999development}. We compute this Maintainability Index (MI) as follows:

\begingroup
\footnotesize
\[
    MI = \max \left[ 0, 100 \frac{171 - 5.2 \ln V - 0.23 G - 16.2 \ln L + 50 \sin \left( \sqrt{2.4 C} \right)}{171} \right]
\]
\endgroup

\noindent where V = Halstead Volume,
    G = Total Cyclomatic Complexity,
    L = Number of Source Lines of Code (SLOC),
    C = Percent of comment lines (converted to radians).

We use Radon \cite{lacchia2023radon}, a widely adopted Python tool, to compute Cyclomatic complexity, Halstead metrics, and the Maintainability Index \cite{malhotra2015python, liawatimena2018django}. Radon requires a parsable Python code to perform these calculations as it parses the actual source code. Our dataset includes 50.6\% (30,113 out of 59,561) of code snippets where both the initial and latest versions were parsable, and the edit distance between the two versions was greater than zero, indicating modifications. We thus analyze these to assess whether the edits lead to a reduction in code complexity.

% \begin{table}[htbp]
% % \vspace{-2mm}
%     \centering
%     \caption{SonarQube Severity Mapping}
%     \label{table:severity-mapping}
%     \resizebox{2in}{!}{%
%     \begin{tabular}{c|c}
%     \toprule
%     \textbf{Severity} & \textbf{Mapped to} \\ \midrule
%     Blocker, Critical & High \\ \midrule
%     Major & Medium \\ \midrule
%     Minor, Info & Low \\  \bottomrule
%     \end{tabular}
%     }
%     % \vspace{-5mm}
% \end{table}

\subsection{Code Vulnerability Assessment (RQ4)}
\label{subsec-mothodology:vulnerability}

Code vulnerabilities (e.g., bugs, code smells, security) are crucial as they affect software reliability, maintainability, performance, and user trust.
We leverage SonarQube \cite{sonarqube} to detect bugs and code smells. It is a widely used static code analysis tool that aids in detecting code issues \cite{marcilio2019static, alfayez2023sonarqube}. It defines 84 rules for detecting bugs and 159 for identifying code smells, tailored explicitly to the Python programming language \cite{sonarquberules}. 
In SonarQube, a bug is a coding error that causes incorrect behavior or crashes that need immediate fixing. A code smell is a bad coding practice that does not break functionality but makes the code harder to maintain and should be refactored for better quality.
% Bugs represent coding mistakes that could lead to runtime errors or unexpected behavior. Code smells indicate suboptimal code structures that hinder maintainability, making the code more complex and challenging to manage. 
% SonarQube assigns severity levels to these issues based on their impact on software quality and maintainability, as follows: \{Blocker, Critical\} $\rightarrow$ High, \{Major, Medium\} $\rightarrow$ Medium, and \{Minor, Info\} $\rightarrow$ Low.
%illustrated in Table \ref{table:severity-mapping}.

We assess code security using Bandit \cite{bandit-security}, a popular static analysis tool designed to detect common security vulnerabilities in Python code \cite{kapustin2023static, peng2019python, lamalva2023python}. Developed by the OpenStack Security Project \cite{openstackbandit}, Bandit analyzes Python code by constructing an Abstract Syntax Tree (AST) and applying a predefined set of security rules. These rules target vulnerabilities such as the use of insecure functions, improper handling of user input, and misconfigured security settings \cite{bandit-security}. By examining the AST, Bandit identifies structural patterns associated with known security risks. Once the analysis is complete, a report detailing detected security issues is generated.
Similar to the previous section (Section \ref{subsec-mothodology:complexity}), we use 30,113 code snippets where the initial and latest versions were parsable for bugs, code smells, and security analysis. In particular, we examine whether code edits result in a reduction of these code issues.

\subsection{Code Optimization Assessment (RQ5)}
\label{subsec-mothodology:optimization}

Code optimization is crucial for enhancing software efficiency and reducing execution time, which directly impacts scalability and performance. To assess the effect of edits on optimization, we analyze the total number of function calls and execution time (milliseconds).
We measure function calls and execution time using \textit{cProfile}, a built-in Python profiler that tracks function calls and records cumulative execution time. 
% For memory usage, we employ \textit{memory\_profiler}, which captures the peak memory consumption of a running code snippet.
% Similar to the previous section (Section \ref{subsec-mothodology:complexity}),
Since this measure requires code execution, we use 7,446 answers where code snippets in the initial and latest versions were executable. Our goal is to determine whether code edits reduce function calls and execution time, thereby enhancing code optimization.

\subsection{Readability Assessment (RQ6)}
\label{subsec-mothodology:readability}

\textit{Text readability} depends on vocabulary complexity, syntactic structure, and presentation styles such as font size, line height, and line length \cite{readabilitywiki}. In our dataset, 69,007 answers underwent text edits where the edit distance of texts between the initial and latest versions is greater than zero. 
To assess whether these edits improve readability, we compute one of the widely used \cite{mondal2023subjectivity, rahman2015insight, ponzanelli2014understanding} standardized readability metrics: Flesch Reading Ease (FRE) \cite{flesch1948new}. This metric assesses text readability based on sentence length, words count, and syllable count as follows:

\smallskip
{\small
$\text{FRE} = 206.835 - 1.015 \times \left(\frac{\text{Total Words}}{\text{Total Sentences}}\right) - 84.6 \times \left(\frac{\text{Total Syllables}}{\text{Total Words}}\right)$
}
\smallskip

\noindent The scale ranges from 0 to 100, where higher scores indicate easier readability.
% To assess whether these edits improve readability, we compute six standardized readability metrics: Automated Readability Index \cite{smith1967automated}, Flesch-Kincaid Grade Level \cite{flesch1948new}, Coleman-Liau Index \cite{coleman1975computer}, SMOG Grade \cite{mc1969smog}, Gunning Fog Index \cite{gunning1952technique}, and Flesch Reading Ease Score \cite{flesch1948new}. These metrics estimate the comprehension difficulty of English texts, each providing a different approximation of the U.S. grade level required for understanding. We first compute the readability score for each text, normalize the values between 0 and 100, and then calculate the average readability score across the five indices. A lower score indicates higher readability, while a higher score suggests greater reading difficulty.

\textit{Code readability} refers to how easily a code snippet can be read and understood by developers who did not author it \cite{ASM-Posnett-2011}. Code readability is strongly related to software maintainability and quality. Thus, the underlying idea is that the more readable the code is, the easier it is to reuse and maintain in the long run. Since poor readability of code costs development time and effort, we analyze whether edits improve the readability of the code snippets added to the answers. 
Specifically, we examine whether there is a noticeable difference in code readability between the initial and latest versions of code snippets. To this end, we analyze 59,561 answers (Section \ref{subsec-mothodology:usability}) containing code snippets in both the initial and latest versions, where the edit distance is greater than zero, to assess whether edits improve code readability.

To compute the code readability, we use the model developed by Buse and Weimer \cite{LAM-Buse-2010}, which is trained on the human perception of readability or understandability. This model extracts textual features from the code that are likely to affect the humans’ perception of readability and predicts a readability score on a scale from zero to one. One indicates that the source code is highly readable, whereas zero indicates poor readability.

\subsection{Environment Setup}
\label{subsec:envSetup}

We use \textit{Visual Studio Code}\footnote{https://code.visualstudio.com} (Version: 1.98.0) to analyze code snippets and compute the quality metrics. Visual Studio Code is a widely used, feature-rich code editor for Python-based research and software development. We use \textit{Python 3.8.8} as Python interpreter\footnote{https://www.python.org} for executing and analyzing code snippets. We use a desktop computer with a 64-bit Windows 10 Operating System (OS) equipped with an Intel(R) Core(TM) i7-7700 processor and 16GB primary memory (i.e., RAM).

% \subsection{AI-Driven Code Quality Judgment (RQ7)}
% \label{subsec-mothodology:ai-driven-code-quality}

\section{Findings and Analysis on Relevance, Code Quality and Readability (RQ1--RQ6)}
\label{sec:findings}

\begin{figure}[!htb]
\centering
\resizebox{2.4in}{!}{%
\begin{tikzpicture}
\begin{axis}[
    boxplot/draw direction=y,
    ymode=normal, % Ensure normal y-axis scale
    scaled y ticks=false, % Disable scientific notation
    yticklabel style={/pgf/number format/fixed}, % Display numbers in fixed-point notation
    % ymin=0, ymax=30000, % Define y-axis limits
    % ytick={0, 5000, 10000, 15000, 20000, 25000, 30000}, % Set y-axis tick values explicitly
    ylabel={Value}, % Set y-axis label
    ylabel={Relevance score},
    xtick={1,2},
    xticklabels={
        Initial Version, 
        Latest Version
    },
    % x tick label style={text width=1.5cm,align=center},
    width=7cm,
    height=7.5cm
]

% Compatibility Issues
\addplot+[
    boxplot prepared={
    median=0.69,
    upper quartile=0.76,
    lower quartile=0.60,
    upper whisker=0.96,
    lower whisker=0.35
    },
] coordinates {};

% Installation and Package Dependency Issues
\addplot+[
    boxplot prepared={
    median=0.71,
    upper quartile=0.77,
    lower quartile=0.63,
    upper whisker=0.96,
    lower whisker=0.40
    },
] coordinates {};

\end{axis}
\end{tikzpicture}
}

\caption{Semantic relevance between questions and the initial and latest versions of answers.}
\label{fig:boxplot-semantic relevance}
\end{figure}
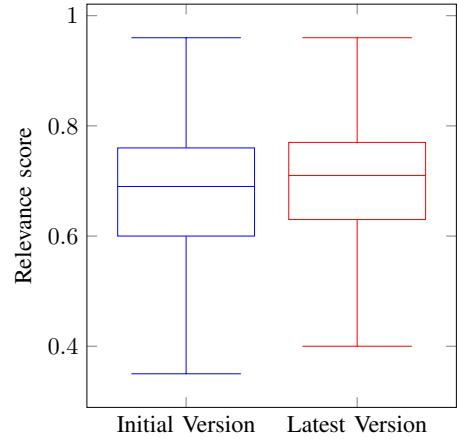

\subsection{Findings on the Impact of Editing on Semantic Relevance and Alignment in Question-Answer Pairs (RQ1)}
\label{subsec:findings-semantic-relevance}

Fig. \ref{fig:boxplot-semantic relevance} shows the boxplots of similarity scores of the initial and latest versions of answers with questions. The median similarity score for the latest versions (0.71) is higher than that of the initial versions (0.69). We then use the \emph{Mann-Whitney-Wilcoxon} test to check whether the score difference is statistically significant. We also measure their effect size using the \emph{Cliff's Delta} test \cite{macbeth2011cliff}). We find statistically significant \emph{p-value} (p-value $\approx$ 0.0 $<$ 0.05) from our analysis. However, the Cliff's Delta effect size is negligible ($|$d$|$ $= 0.08$, confidence level $95$). These results suggest that edits slightly improve answer coherence with questions but have minimal practical impact.

To gain deeper insights, we analyze how often edits improve or reduce the relevance of answers to their corresponding questions. According to our analysis, in 53.3\% of cases (50,616 out of 94,995), edits enhance relevance, whereas in 38.1\% of cases (36,175 out of 94,995), edits reduce relevance. The Mann-Whitney-Wilcoxon test confirms statistical significance (p-value $\approx$ 0.0 $<$ 0.05) for both cases. However, the effect size differs: when edits improve relevance, the effect size is small ($|$d$|$ $= 0.23$), while for cases where relevance decreases, the effect size is negligible ($|$d$|$ $= 0.13$).

\begin{findingbox}
\leftskip 10pt \rightskip 5pt \textbf{Summary of RQ\textsubscript{1}}:
Edits generally improve answer relevance (53.3\% of cases), but a significant portion (38.1\%) reduces alignment with questions. The small effect size ($|$d$|$ $= 0.23$) for relevance-enhancing edits suggests modest but consistent improvements, whereas the negligible effect size ($|$d$|$ $= 0.13$) for relevance-reducing edits indicates minimal semantic disruption.
\end{findingbox}

\begin{figure}[!htb]
\centering
    \pgfplotstableread{
    		1	59.2   56.5     35241   33657 
    		2	17.4   20.0     10383    11883
    }\datatable
    \subfloat[Percentage]{
    \label{fig:reproducibility-status-ratio}
   	  \resizebox{1.7in}{!}{%
      \begin{tikzpicture}
        	\begin{axis}[
        	xtick=data,
        	xticklabels={Parsability, Executability},
        	enlarge y limits=false,
        	enlarge y limits=false,
        	enlarge x limits=0.60,
        %	nodes near coords,
        	ymin=0,ymax=100,
        	ybar,
        	bar width=0.6cm,
        	width=2.5in,
        	height = 2.5in,
        	ytick={0,20,...,100},
            yticklabels={0\%,20\%,40\%,60\%,80\%,100\%,},
        	ymajorgrids=false,
        %	xminorgrids=true,
        % 	yticklabel style={font=\small},
        % 	xticklabel style={font=\footnotesize, /pgf/number format/fixed},	
        	major x tick style = {opacity=0},
        	minor x tick num = 1,    
        	minor tick length=1ex,
        	legend style={
        % 	at={(0.5,-0.20)},
        % 	font=\small,
        % 	legend pos = outer north east,
         	legend pos=north east,
        %   anchor=west,
        % 	cells={align=left},
        	legend cell align=left
            },
            nodes near coords style={rotate=90,  anchor=west}, %font=\footnotesize
        	nodes near coords =\pgfmathprintnumber{\pgfplotspointmeta}\%
        	%nodes near coords*={\pgfmathprintnumber[precision=2]\pgfplotspointmeta \%}
        	]
        	\addplot[draw=black!80, fill=black!0] table[x index=0,y index=1] \datatable;
        	\addplot[draw=black!80, fill=black!50] table[x index=0,y index=2] \datatable;

            \legend	{Initial,
            		 Latest
            		 }
        	\end{axis}
    	\end{tikzpicture}
    	}
    	}
      \subfloat[Count]{
      \label{fig:reproducibility-status-count}
   	  \resizebox{1.7in}{!}{%
      \begin{tikzpicture}
        	\begin{axis}[
        	xtick=data,
        	xticklabels={Parsability, Executability},
        	enlarge y limits=false,
        	enlarge x limits=0.60,
        %	nodes near coords,
        	ymin=0,ymax=50000,
        	ybar,
        	bar width=0.6cm,
        	width=2.5in,
        	height = 2.55in,
        	ytick={0,5000,...,50000},
                scaled y ticks=false,
        	ymajorgrids=false,
        %	xminorgrids=true,
        % 	yticklabel style={font=\small},
        % 	xticklabel style={font=\footnotesize, /pgf/number format/fixed},	
        	major x tick style = {opacity=0},
        	minor x tick num = 1,    
        	minor tick length=1ex,
        	legend style={
        % 	at={(0.5,-0.20)},
        % 	font=\small,
        % 	legend pos = outer north east,
         	legend pos=north east,
        %   anchor=west,
        % 	cells={align=left},
        	legend cell align=left
            },
            nodes near coords style={rotate=90,  anchor=west}, %font=\footnotesize
        	nodes near coords =\pgfmathprintnumber{\pgfplotspointmeta}%\%
        	%nodes near coords*={\pgfmathprintnumber[precision=2]\pgfplotspointmeta \%}
        	]
        	\addplot[draw=black!80, fill=black!0] table[x index=0,y index=3] \datatable;
        	\addplot[draw=black!80, fill=black!50] table[x index=0,y index=4] \datatable;

            \legend{Initial,
            		 Latest
            		 }
        	\end{axis}
    	\end{tikzpicture}
    	}
    	}
    %\vspace{-3mm}
\caption{Usability comparison between initial and latest versions of answer code snippets.}
%\vspace{-1mm}
\label{fig:parsability-summary}
\end{figure}
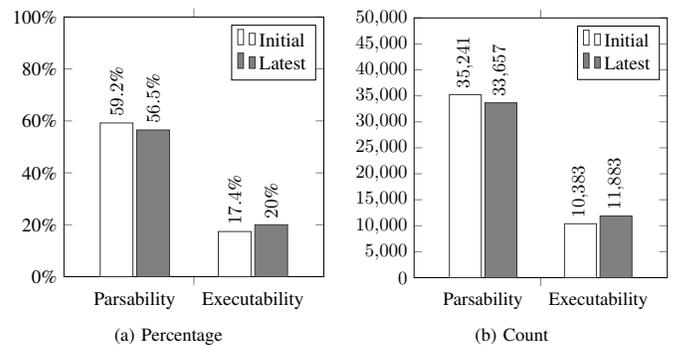

\begin{table}[!t]
\centering
    \caption{Cyclomatic complexity comparison between the initial and latest versions of answer code snippets (\textbf{M} = Mean, \textbf{Med} = Median, \textbf{IV} = Initial Version, \textbf{LV} = Latest Version).}
    \label{table:cyclomatic-complexity}
    \resizebox{3.4in}{!}{%
    \begin{tabular}{lcccccc}
    \toprule
    
    \textbf{Version} & \textbf{M} & \textbf{Med} & \textbf{LV\textless{}IV} & \textbf{LV\textgreater{}IV} & \textbf{p-value} & \textbf{$|$d$|$} \\ \midrule
    
    \textbf{Initial} & 0.7 & 0.0 &  \multirow{2}{*}{6.3\%} & \multirow{2}{*}{32.3\%} & \multirow{2}{*}{$\approx$ 0.0} & \multirow{2}{*}{0.07 (Neg.)} \\ \cmidrule{1-3}
    \textbf{Latest} & 1.1 & 0.0 &  &  &  &  \\ \bottomrule

    \end{tabular}
    }
\end{table}

\begin{table*}[]
\centering
    \caption{Halstead complexity comparison between the initial and latest versions of answer code snippets (\textbf{M} = Mean, \textbf{Med} = Median, \textbf{IV} = Initial Version, \textbf{LV} = Latest Version).}
    \label{table:halstead-complexity}
    \resizebox{4.5in}{!}{%
    \begin{tabular}{llcccccc}
    \toprule
    
    \textbf{Measures} & \textbf{Version} & \textbf{M} & \textbf{Med} & \textbf{LV\textless{}IV} & \textbf{LV\textgreater{}IV} & \textbf{p-value} & \textbf{$|$d$|$} \\ \midrule
    
    \multirow{2}{*}{\textbf{Volume}} & \textbf{Initial} & 10.0 & 4.8 &  \multirow{2}{*}{6.6\%} & \multirow{2}{*}{30.9\%} & \multirow{2}{*}{$\approx$ 0.0} & \multirow{2}{*}{0.13 (Neg.)} \\ \cmidrule{2-4}
    & \textbf{Latest} & 15.8 & 4.8 &  &  &  &  \\ \midrule

    \multirow{2}{*}{\textbf{Difficulty}} & \textbf{Initial} & 0.6 & 0.5 &  \multirow{2}{*}{7.0\%} & \multirow{2}{*}{28.0\%} & \multirow{2}{*}{$\approx$ 0.0} & \multirow{2}{*}{0.13 (Neg.)} \\ \cmidrule{2-4}
    & \textbf{Latest} & 0.8 & 0.5 &  &  &  &  \\ \midrule

    \multirow{2}{*}{\textbf{Effort}} & \textbf{Initial} & 7.8 & 2.4 &  \multirow{2}{*}{6.7\%} & \multirow{2}{*}{30.9\%} & \multirow{2}{*}{$\approx$ 0.0} & \multirow{2}{*}{0.14 (Neg.)} \\ \cmidrule{2-4}
    & \textbf{Latest} & 16.1 & 2.4 &  &  &  &  \\ \bottomrule

    \end{tabular}
    }
\end{table*}

\subsection{Findings on the Role of Edits in Enhancing Code Parsability and Executability (RQ2)}
\label{subsec:findings-usability}

Fig. \ref{fig:parsability-summary} shows a comparative analysis of parsability and executability between the initial and latest versions of answer code snippets. In the initial versions, 59.2\% (35,241 out of 59,562) of code snippets were parsable, whereas in the latest versions, this slightly declined to 56.5\% (33,657 out of 59,562). Notably, 14.6\% (5,128 out of 35,241) of initially parsable snippets became non-parsable. Such findings suggest that several edits introduced syntax errors or structural inconsistencies. Similarly, 14.6\% (3,544 out of 24,321) of initially non-parsable snippets became parsable, which suggests that some edits effectively corrected or refined the code structure. Surprisingly, 14.7\% (1,527 out of 10,383) of initially executable snippets became non-parsable, demonstrating that specific edits disrupted the code structure, causing execution failures. This is likely caused by missing dependencies, incorrect modifications, or compatibility issues introduced during editing.

On the other hand, the overall executability of code snippets improved, increasing from 17.4\% (10,383 out of 59,562) in the initial version to 20.0\% (11,883 out of 59,562) in the latest version. However, 9.0\% (4,437 out of 49,179) of initially non-executable answers became executable, which indicates that some edits successfully addressed errors and improved code correctness. Interestingly, 18.1\% (4,390 out of 24,321) of initially non-parsable answers became executable, which demonstrates that certain edits not only resolved syntax issues but also introduced necessary modifications that enabled successful execution.

\begin{findingbox}
\leftskip 10pt \rightskip 5pt \textbf{Summary of RQ\textsubscript{2}}:
Edits on answer code snippets show mixed effects on usability, with executability improving (17.4\% → 20.0\%) but parsability declining (59.2\% → 56.5\%). While 18.1\% of non-parsable snippets became executable, 14.7\% of executable snippets became non-parsable. That is, some edits fix broken code while others introduce structural issues. 
These findings highlight the inconsistent impact of edits on code snippet usability and suggest clear guidelines and automated validation.
\end{findingbox}

\subsection{Findings on Code Complexity Reduction and Maintainability Through Edits (RQ3)}
\label{subsec:findings-code-complexity}

\textbf{Cyclomatic Complexity.} Table \ref{table:cyclomatic-complexity} presents a comparative analysis of cyclomatic complexity between the initial (IV) and latest (LV) versions of answer code snippets. The majority of code snippets in both versions exhibit simple procedures with minimal risk, with 62.9\% having a complexity of zero. While the mean complexity slightly increases in the latest versions (0.7 to 1.1), the median remains unchanged at zero, which suggests that most edits do not introduce additional branching logic. A statistically significant shift (p-value $\approx$ 0.0 $<$ 0.05) is observed using the Mann-Whitney-Wilcoxon test with negligible effect ($|$d$|$ $= 0.07$). Please note that we remove outliers to ensure a more accurate analysis and avoid skewed results.

A further breakdown of complexity distribution reveals that 6.3\% (607 out of 9,585) of snippets decrease in complexity, while 32.3\% (3,097 out of 9,585) increase. In both versions (with non-zero cyclomatic complexity), most snippets fall into the low-risk category (complexity 1--10). In particular, 95.0\% (9,107 out of 9,585) of initial snippets are low-risk, while in the latest version, this percentage decreases to 91.6\% (10,119 out of 11,048), indicating a slight shift toward higher complexity in edited code. However, moderate-risk (complexity 11–-20) increased from 4.4\% to 7.1\%, and high-risk (complexity 21–50) from 58 sample count to 138. Critically, 7 code snippets found in the latest version of answers now exceed a complexity of 50, causing them to be untestable. Notably, 384 snippets shift from low to moderate risk and 114 from low to high risk. These findings highlight that edits do not necessarily reduce complexity. Instead, they sometimes increase it, likely due to the addition of more code rather than structural simplification.

\textbf{Halstead complexity.}
Table \ref{table:halstead-complexity} compares the Halstead complexity between the initial and latest versions of answer code snippets.
Across volume, difficulty, and effort, the mean (M) values increase in the latest version, which indicates a general rise in complexity, while the median (Med) values remain stable. The mean volume increases from 10.0 (initial) to 15.8 (latest), but the median stays at 4.8. Mean difficulty rises slightly from 0.6 to 0.8, while the median stays at 0.5, which shows minimal impact on most snippets. The mean effort almost doubles from 7.8 to 16.1, yet the median remains at 2.4, which confirms that complexity increases in only a subset of cases rather than universally. The Mann-Whitney-Wilcoxon test confirms statistical significance (p-value $\approx$ 0.0 $<$ 0.05), though the effect sizes remain negligible ($|$d$|$ $ \approx 0.13-0.14$), suggesting that while these changes are frequent, their practical impact on individual snippets is limited.

A more detailed examination shows that 49.2\% of snippets (9,301 out of 18,896) experience an increase in volume, while only 10.5\% (1,980 out of 18,896) see a decrease, which indicates a general expansion of code size. Similarly, the difficulty increases in 44.5\% of cases (8,418 out of 18,896), whereas only 11.1\% (2,097 snippets) experience reductions, highlighting a growing complexity in operations. The effort, which encapsulates both volume and difficulty, follows the same trend, with 49.2\% of snippets (9,302 out of 18,896) requiring more effort and only 10.5\% (1,985) requiring less. These trends suggest that edits often introduce additional logic and operations and increase complexity. While such changes may improve functionality, they can also raise the cognitive burden for developers.

\begin{table}[!htb]
\centering
    \caption{Maintainability index comparison between the initial and latest versions of answer code snippets (\textbf{M} = Mean, \textbf{Med} = Median, \textbf{IV} = Initial Version, \textbf{LV} = Latest Version).}
    \label{table:maintainability-index-complexity}
    \resizebox{3.4in}{!}{%
    \begin{tabular}{lcccccc}
    \toprule
    
    \textbf{Version} & \textbf{M} & \textbf{Med} & \textbf{LV\textless{}IV} & \textbf{LV\textgreater{}IV} & \textbf{p-value} & \textbf{$|$d$|$} \\ \midrule
    
    \textbf{Initial} & 86.6 & 90.7 &  \multirow{2}{*}{37.5\%} & \multirow{2}{*}{15.4\%} & \multirow{2}{*}{$\approx$ 0.0} & \multirow{2}{*}{0.09 (Neg.)} \\ \cmidrule{1-3}
    \textbf{Latest} & 84.2 & 85.4 &  &  &  &  \\ \bottomrule

    \end{tabular}
    }
\end{table}

\begin{table*}[!htb]
\centering
    \caption{Code vulnerability (bug, smell, and security vulnerabilities) comparison between the initial and latest versions of answer code snippets (\textbf{M} = Mean, \textbf{Med} = Median, \textbf{IV} = Initial Version, \textbf{LV} = Latest Version).}
    \label{table:code-vulnerability}
    \resizebox{6.5in}{!}{%
    \begin{tabular}{llcccccccc}
    \toprule
    
    \textbf{Measures} & \textbf{Version} & \textbf{M} & \textbf{Med} & \textbf{LV\textless{}IV} & \textbf{LV\textgreater{}IV} & \textbf{LV=IV} & \textbf{IV=0, LV\textgreater{}0} & \textbf{p-value} & \textbf{$|$d$|$} \\ \midrule
    
    \multirow{2}{*}{\textbf{Bugs}} & \textbf{Initial} & 0.3 & 0.0 &  
    \multirow{2}{*}{\begin{tabular}[c]{@{}c@{}}23.1\% \\ (990/4,292)\end{tabular}} & 
    \multirow{2}{*}{\begin{tabular}[c]{@{}c@{}}17.9\% \\ (770/4,292)\end{tabular}} &  \multirow{2}{*}{\begin{tabular}[c]{@{}c@{}}59.0\% \\ (2,531/4292)\end{tabular}} & 
    \multirow{2}{*}{\begin{tabular}[c]{@{}c@{}}6.6\% \\ (1,711/25,822)\end{tabular}} & 
    \multirow{2}{*}{$\approx$ 0.0} & \multirow{2}{*}{0.03 (Neg.)} \\ \cmidrule{2-4}
    & \textbf{Latest} & 0.5 & 0.0 &  &  &  & &  &  \\ \midrule

    \multirow{2}{*}{\textbf{Smells}} & \textbf{Initial} & 0.3 & 0.0 &  
    \multirow{2}{*}{\begin{tabular}[c]{@{}c@{}}27.5\% \\ (1,227/4,461)\end{tabular}} & 
    \multirow{2}{*}{\begin{tabular}[c]{@{}c@{}}16.3\% \\ (729/4,461)\end{tabular}} &  \multirow{2}{*}{\begin{tabular}[c]{@{}c@{}}56.1\% \\ (2,504/4,461)\end{tabular}} & 
    \multirow{2}{*}{\begin{tabular}[c]{@{}c@{}}7.9\% \\ (2,033/25,653)\end{tabular}} & 
    \multirow{2}{*}{$\approx$ 0.0} & \multirow{2}{*}{0.04 (Neg.)} \\ \cmidrule{2-4}
    & \textbf{Latest} & 0.4 & 0.0 &  &  &  & &  &  \\ \midrule
    
    \multirow{2}{*}{\textbf{Security}} & \textbf{Initial} & 0.1 & 0.0 &  
    \multirow{2}{*}{\begin{tabular}[c]{@{}c@{}}7.5\% \\ (162/2,154)\end{tabular}} & 
    \multirow{2}{*}{\begin{tabular}[c]{@{}c@{}}20.5\% \\ (441/2,154)\end{tabular}} &  \multirow{2}{*}{\begin{tabular}[c]{@{}c@{}}72.0\% \\ (1,550/2,154)\end{tabular}} & 
    \multirow{2}{*}{\begin{tabular}[c]{@{}c@{}}1.4\% \\ (401/27,960)\end{tabular}} & 
    \multirow{2}{*}{$\approx$ 0.0} & \multirow{2}{*}{0.01 (Neg.)} \\ \cmidrule{2-4}
    & \textbf{Latest} & 0.2 & 0.0 &  &  &  & &  &  \\ \bottomrule

    \end{tabular}
    }
\end{table*}

\textbf{Maintainability Index.}
Table \ref{table:maintainability-index-complexity} compares the Maintainability Index between the initial and latest versions of answer code snippets. The overall maintainability remains good, though the mean (M) decreases from 86.6 to 84.2, and the median (Med) drops from 90.7 to 85.4. A statistically significant shift (p-value $\approx$ 0.0 $<$ 0.05) is observed, though the effect size is negligible ($|$d$|$ $ = 0.09$). A deeper breakdown shows that 37.5\% of snippets (11,276 out of 30,113) experience a decline in maintainability, whereas 15.4\% (4,638 snippets) improve.

\begin{findingbox}
\leftskip 10pt \rightskip 5pt \textbf{Summary of RQ\textsubscript{3}}:
The analysis of answer code snippets shows that edits generally increase both logical and cognitive complexity, leading to a decmidrule in maintainability. Cyclomatic complexity rises (mean: 0.7 → 1.1), with 32.3\% of snippets becoming more complex, while Halstead metrics confirm increases in volume (10.0 → 15.8), difficulty (0.6 → 0.8), and effort (7.8 → 16.1) on average, indicating a growing cognitive load. As a result, 37.5\% of snippets see reduced maintainability (MI: 86.6 → 84.2). However, edits may append additional code to enhance functionality, which often introduces more logic instead of simplifying structures, making the code harder to maintain over time.
\end{findingbox}

\begin{table*}[!htb]
\centering
    \caption{Comparison of Common Weakness Enumeration (CWE) codes between the initial version (IV) and latest version (LV), along with their associated issue codes (Detailed issue descriptions can be found in our online appendix \cite{replicationPackage}).}
    \label{table:security-details}
    \resizebox{7in}{!}{%
    \begin{tabular}{p{4.8cm}p{4.5cm}p{4.7cm}}
    \toprule
    \textbf{CWE Code With Details} & \textbf{Count (IV) with Issue Codes} & \textbf{Count (LV) with Issue Codes} \\ \midrule

    \textbf{CWE-330:} Use of Insufficiently Random Values  & 1,511 \{B311 (1,511)\} & 1,967 \{B311 (1,967)\} \\ \midrule

    \textbf{CWE-78:} Improper Neutralization of Special Elements used in an OS Command (`OS Command Injection')    & 1,078 \{B102 (72), B307 (197), B404 (154), B602 (59), B603 (202), B604 (9), B605 (147), B606 (22), B607 (216)\} & 1,467 \{B102 (109), B307 (216), B404 (224), B602 (94), B603 (274), B604 (8), B605 (201), B606 (27), B607 (314)\} \\ \midrule

    \textbf{CWE-703:} Improper Check or Handling of Exceptional Conditions & 458 \{B101 (372), B110 (81), B112 (5)\} & 678 \{B101 (588), B110 (79), B112 (11)\} \\ \midrule

    \textbf{CWE-400:} Uncontrolled Resource Consumption  & 223 \{B113 (223)\} & 282 \{B113 (282)\} \\ \midrule

    \textbf{CWE-502:} Deserialization of Untrusted Data  & 82 \{B301 (46), B403 (36)\} & 101 \{B301 (55), B403 (46)\} \\ \midrule

    \textbf{CWE-377:} Insecure Temporary File  & 71 \{B108 (71)\} & 90 \{B108 (90)\} \\ \midrule

    \textbf{CWE-259:} Use of Hard-coded Password  & 63 \{B105 (58), B106 (5), B107 (0)\} & 91 \{B105 (82), B106 (8), B107 (1)\} \\ \midrule

    \textbf{CWE-22:} Improper Limitation of a Pathname to a Restricted Directory  (`Path Traversal')  & 53 \{B202 (1), B310 (52)\} & 59 \{B202 (1), B310 (58)\} \\ \midrule

    \textbf{CWE-20:} Improper Input Validation  & 42 \{B314 (10), B320 (2), B405 (14), B410 (12), B411 (1), B506 (3)\} & 60 \{B314 (13), B320 (6), B405 (18), B410 (19), B411 (1), B506 (3)\} \\ \midrule

    \textbf{CWE-89:} Improper Neutralization of Special Elements used in an SQL Command  (`SQL Injection')  & 27 \{B608 (27), B610 (0)\} & 35 \{B608 (34), B610 (1)\} \\ \midrule

    \textbf{CWE-327:} Use of a Broken or Risky Cryptographic Algorithm  & 12 \{B303 (9), B324 (1), B413 (1), B504 (1)\} & 14 \{B303 (13), B324 (1), B413 (0), B504 (0)\} \\ \midrule

    \textbf{CWE-319:} Cleartext Transmission of Sensitive Information  & 9 \{B312 (1), B321 (4), B401 (1), B402 (3)\} & 11 \{B312 (2), B321 (4), B401 (2), B402 (3)\} \\ \midrule

    \textbf{CWE-295:} Improper Certificate Validation  & 6 \{B323 (0), B501 (6)\} & 9 \{B323 (1), B501 (8)\} \\ \midrule

    \textbf{CWE-732:} Incorrect Permission Assignment for Critical Resource    & 3 \{B103(3)\} & 2 \{B103(2)\} \\ \midrule

    \textbf{CWE-605:} Multiple Binds to the Same Port  & 1 \{B104(1)\} & 0 \{B104(0)\} \\ \bottomrule

    \end{tabular}
    }
\end{table*}

\subsection{Findings on the Effect of Edits in Reducing Code Vulnerabilities, Bugs, and Code Smells (RQ4)}
\label{subsec:findings-code-vulnerabilities}

In RQ3 (Section \ref{subsec:findings-code-complexity}), we found that editing often increases code complexity, possibly due to adding more code. However, in this RQ, we examine whether editing reduces bugs and code smells or introduces new ones.
Table \ref{table:code-vulnerability} presents a comparative analysis of bugs, smells, and security vulnerabilities between the initial version (IV) and the latest version (LV) of answer code snippets.

\textbf{Bugs.}
The mean bug count rises from 0.3 to 0.5, while the median remains 0.0 because most snippets are bug-free. In the initial version, 85.7\% (25,822 out of 30,144) had no bugs, dropping slightly to 82.9\% (24,988 out of 30,144) in the latest version. A statistically significant shift (p-value $\approx$ 0.0 $<$ 0.05) is found using the Mann-Whitney-Wilcoxon test, though the effect size is negligible ($|$d$|$ $= 0.03$). To ensure accuracy, we remove outliers to prevent skewed results and misinterpretations.

A further breakdown of bug occurrence reveals that 23.1\% (990 out of 4,292) of edited snippets resolved pre-existing bugs (78.9\% fully resolved and 21.1\% partially resolved). In comparison, 17.9\% (770 out of 4,292) introduced new ones with existing bugs. The majority, 59.0\% (2,531 out of 4,292), retained the same number of bugs, which indicates that most of the time, edits overlook these bugs. Surprisingly, editing introduced new bugs in 6.6\% (1,711 out of 25,822) of previously bug-free snippets. These findings indicate that while some edits improve code quality by fixing bugs, most overlook existing ones, and a portion inadvertently introduces new bugs.

\textbf{Code Smells.}
The mean code smell count increases from 0.3 to 0.4, while the median remains 0.0 because most snippets are free of smells. In the initial version, 85.2\% (25,653 out of 30,144) had no detectable smells, decreasing slightly to 81.6\% (24,611 out of 30,144) in the latest version. A statistically significant shift (p-value $\approx$ 0.0 $<$ 0.05) is found using the Mann-Whitney-Wilcoxon test, though the effect size remains negligible ($|$d$|$ $= 0.04$). 

A further breakdown of smell occurrences shows that 27.5\% (1,227 out of 4,461) of edited snippets eliminated pre-existing smells, with 80.8\% fully resolved and 19.2\% partially resolved. In contrast, 16.3\% (729 out of 4,461) introduced new smells alongside existing ones. The majority, 56.1\% (2,504 out of 4,461), retained the same number of smells, suggesting that most edits do not address existing smells. Notably, 7.9\% (2,033 out of 25,653) of previously smell-free snippets acquired smells due to edits. Similar to bugs, while some edits improve code quality by removing smells, most do not address existing issues, and some unintentionally introduce new smells.

\textbf{Security Vulnerabilities.}
The mean security vulnerability count increases from 0.1 to 0.2. However, similar to bugs and smells, the median remains 0.0 because most snippets remain free of security issues. Our further analysis shows that 7.5\% (162 out of 2,154) of edited snippets resolved pre-existing vulnerabilities (70.4\% fully resolved and 29.6\% partially resolved), while 20.5\% (441 out of 2,154) introduced additional ones. The majority, 72.0\% (1,550 out of 2,154), retained the same number of security vulnerabilities, indicating that most edits overlook these issues. Notably, 1.4\% (401 out of 27,960) of previously secure snippets introduced security vulnerabilities through edits. Similar to bugs and smells, while some edits improve code security, most do not address existing vulnerabilities, and some introduce new ones. These findings also align with existing studies \cite{verdi2020empirical, zhang2021study}, which investigated C/C++ security vulnerabilities and found that they are rarely addressed, with most edits ignoring security issues.

Table \ref{table:security-details} further shows the breakdown of Common Weakness Enumeration (CWE) and associated issues. We see that the most frequent security vulnerabilities in both versions are OS command injection (CWE-78), use of insufficiently random values (CWE-330), and improper handling of exceptional conditions (CWE-703). CWE-78 remains the most prevalent, rising from 1,078 to 1,467 instances, while CWE-330 increased from 1,511 to 1,967, indicating persistent weaknesses in randomness for security-critical operations. Similarly, CWE-703 rose from 458 to 678, suggesting error handling remains a common issue in modified code. Notably, high-risk vulnerabilities like CWE-89 (SQL injection) and CWE-400 (uncontrolled resource consumption) also increased through code edits. These findings underscore the need for stronger security awareness in code edits, as edits often fail to address, and sometimes worsen, critical vulnerabilities.

\begin{findingbox}
\leftskip 10pt \rightskip 5pt \textbf{Summary of RQ\textsubscript{4}}:
The analysis of answer code snippets reveals that edits often fail to resolve existing code issues and sometimes introduce new ones. Bug count rises (mean: 0.3 → 0.5), with 17.9\% of edits introducing bugs, while 59.0\% overlook existing ones. Similarly, code smells increase (0.3 → 0.4 mean), with 7.9\% of previously clean snippets acquiring smells. Security vulnerabilities also persist (mean: 0.1 → 0.2), with 20.5\% of edits introducing new risks, particularly in OS command injection (CWE-78), weak randomness (CWE-330), and exception handling (CWE-703).
\end{findingbox}

\begin{figure}[!htb]
    \centering
    
    \subfloat[Total function calls \label{fig:boxplot-function-call}]
    {
        \resizebox{1.65in}{!}{%
            \begin{tikzpicture}
            \begin{axis}[
                boxplot/draw direction=y,
                ymode=normal, 
                scaled y ticks=false,
                yticklabel style={/pgf/number format/fixed}, 
                ylabel={Count},
                xtick={1,2},
                xticklabels={IV, LV},
                width=5cm,
                height=6.3cm
            ]

            % Initial Version
            \addplot+[
                boxplot prepared={
                median=6,
                upper quartile=10,
                lower quartile=5,
                upper whisker=35,
                lower whisker=0
                },
            ] coordinates {};

            % Latest Version
            \addplot+[
                boxplot prepared={
                median=7,
                upper quartile=14,
                lower quartile=4,
                upper whisker=61,
                lower whisker=0
                },
            ] coordinates {};

            \end{axis}
            \end{tikzpicture}
        }
    }
    \subfloat[Cummulative execution time. \label{fig:boxplot-cummulative-time}]
    {
        \resizebox{1.65in}{!}{%
            \begin{tikzpicture}
            \begin{axis}[
                boxplot/draw direction=y,
                ymode=normal, 
                scaled y ticks=false,
                yticklabel style={/pgf/number format/fixed}, 
                ylabel={Times (in milliseconds)},
                xtick={1,2},
                xticklabels={IV, LV},
                width=5cm,
                height=6.5cm
            ]

            \addplot+[
                boxplot prepared={
                median=0.23,
                upper quartile=0.43,
                lower quartile=0.13,
                upper whisker=1.23,
                lower whisker=0
                },
            ] coordinates {};

            \addplot+[
                boxplot prepared={
                median=0.24,
                upper quartile=0.52,
                lower quartile=0.12,
                upper whisker=1.65,
                lower whisker=0
                },
            ] coordinates {};

            \end{axis}
            \end{tikzpicture}
        }
    }

    \caption{Box plot comparing total function calls and cumulative execution times between the initial and latest versions of code snippets (\textbf{IV} = Initial Version, \textbf{LV} = Latest Version).}
    \label{fig:boxplot-code-optimization}
\end{figure}
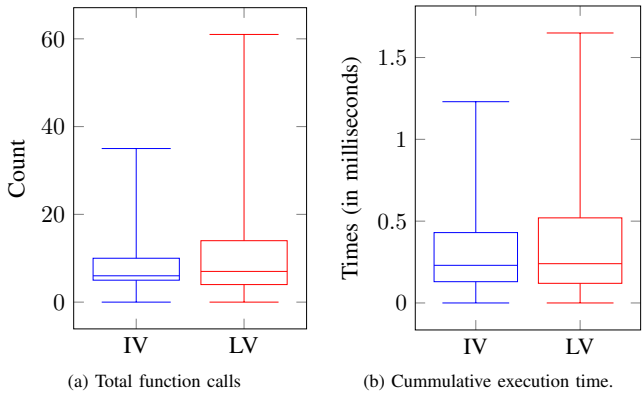

\subsection{Findings on the Impact of Editing on Code Optimization and Performance Enhancement (RQ5)}
\label{subsec:findings-code-optimization}

Fig. \ref{fig:boxplot-function-call} presents a box plot comparison of total function calls between the initial and latest versions of the code snippets. The mean function calls increased from 8.6 to 11.4, while the median increased slightly from 6 to 7. We find statistical significant difference (p-value $=$ 0.003 $<$ 0.05) between the function calls using Mann-Whitney-Wilcoxon test, though the effect size is negligible ($|$d$|$ $= 0.03$). A deeper analysis shows that 49.0\% (3,648 out of 7,447) of edits reduced function calls, whereas 35.2\% (2,621 out of 7,447) increased function calls. Despite more reductions (49.0\% cases), the mean and median increased as increments in function calls were larger than reductions. This suggests that while some edits optimize function usage, others might add complexity through new functions and refactored logic.

Fig. \ref{fig:boxplot-cummulative-time} presents a box plot comparison of cumulative execution times between the initial and latest versions of code snippets. The mean execution time increased from 0.3 to 0.4, while the median remained unchanged at 0.2. A statistically significant difference (p-value $=$ 0.0002 $<$ 0.05) was found using the Mann-Whitney-Wilcoxon test, though the effect size remains negligible ($|$d$|$ $= 0.04$). Further analysis shows that 51.0\% (3,797 out of 7,447) of edits reduced execution time, whereas 45.3\% (3,372 out of 7,447) increased it.
While some edits improve time efficiency, others increase computational overhead (e.g., higher function calls or more complex logic).

\begin{findingbox}
\leftskip 10pt \rightskip 5pt \textbf{Summary of RQ\textsubscript{5}}:
Our analysis shows that function calls and execution times often increase through edits. Function calls rise (mean: 8.6 → 11.4, median: 6 → 7) despite 49.0\% of edits reducing calls, as 35.2\% increase them with larger increments. Similarly, execution time increases (mean: 0.3 → 0.4, median unchanged at 0.2) even though 51.0\% of edits reduce time, as 45.3\% increase often due to higher function calls or more complex logic.
\end{findingbox}

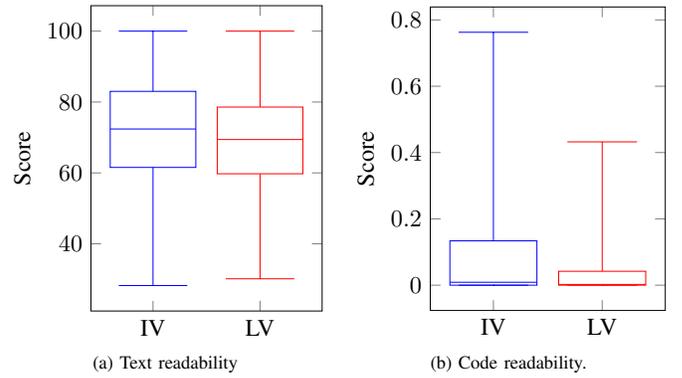
\begin{figure}[!htb]
    \centering
    
    \subfloat[Text readability \label{fig:boxplot-text-readability}]
    {
        \resizebox{1.7in}{!}{%
            \begin{tikzpicture}
            \begin{axis}[
                boxplot/draw direction=y,
                ymode=normal, 
                scaled y ticks=false,
                yticklabel style={/pgf/number format/fixed}, 
                ylabel={Score},
                xtick={1,2},
                xticklabels={IV, LV},
                width=5cm,
                height=6.1cm
            ]

            % Initial Version
            \addplot+[
                boxplot prepared={
                median=72.39,
                upper quartile=83.01,
                lower quartile=61.55,
                upper whisker=100,
                lower whisker=28.22
                },
            ] coordinates {};

            % Latest Version
            \addplot+[
                boxplot prepared={
                median=69.42,
                upper quartile=78.58,
                lower quartile=59.76,
                upper whisker=100,
                lower whisker=30.13
                },
            ] coordinates {};

            \end{axis}
            \end{tikzpicture}
        }
    }
    \subfloat[Code readability. \label{fig:boxplot-code-readability}]
    {
        \resizebox{1.7in}{!}{%
            \begin{tikzpicture}
            \begin{axis}[
                boxplot/draw direction=y,
                ymode=normal, 
                scaled y ticks=false,
                yticklabel style={/pgf/number format/fixed}, 
                ylabel={Score},
                xtick={1,2},
                xticklabels={IV, LV},
                width=5cm,
                height=6cm
            ]

            % Metric A
            \addplot+[
                boxplot prepared={
                median=0.008810038,
                upper quartile=0.1337154 ,
                lower quartile=0,
                upper whisker=0.7627382,
                lower whisker=0
                },
            ] coordinates {};

            % Metric B
            \addplot+[
                boxplot prepared={
                median=0.002285453,
                upper quartile=0.04196553,
                lower quartile=0,
                upper whisker=0.4322232,
                lower whisker=0
                },
            ] coordinates {};

            \end{axis}
            \end{tikzpicture}
        }
    }

    \caption{Box plot comparing text and code readability between the initial and latest versions of code snippets.}
    \label{fig:boxplot-semantic-relevance}
\end{figure}

\subsection{Findings on Improvements in Readability of Stack Overflow Answers Due to Edits (RQ6)}
\label{subsec:findings-readability}

Fig. \ref{fig:boxplot-text-readability} presents a box plot comparison of text readability between the initial and latest versions of textual descriptions. Overall, according to the readability score, the text is fairly easy to read and understand. However, the mean readability score decreased from 72.0 to 69.1, while the median decreased from 72.39 to 69.2. We then used the Mann-Whitney-Wilcoxon test to see whether the difference was statistically significant and found a significant p-value (p-value $\approx$ 0.0 $<$ 0.05). However, the effect size remains negligible. This slight decline in readability suggests that edits may introduce longer or more complex text, making descriptions slightly harder to read while maintaining an overall easy-to-understand level.

Fig. \ref{fig:boxplot-text-readability} compares the code readability between the initial and latest versions of the textual description. 
According to our analysis, the readability of the code snippets found in either the initial or latest version of answers of SO is very poor. For example, a significant fraction of 22.3\% (13,277 out of 59,562) of the code snippets from the initial version have a readability score of zero. Such a statistic is 26.4\% (15,704 out of 59,562) for the latest code versions. Only about 18.9\% of initial code snippets and 14.3\% of latest code snippets have medium or above (i.e., score $\geq 0.5$) readability scores. 
These findings align with Mondal et al. \cite{mondal2023subjectivity}, who found that most code snippets are incomplete and lack proper structure, indentation, and naming conventions, contributing to poor readability. 
A further breakdown shows that 30.0\% of edits improved readability, while 49.7\% worsened it.
Although the \emph{p-values} from \emph{Mann-Whitney-Wilcoxon} statistical test is significant (i.e., \emph{p-values} \textless 0.05), the \emph{d-values} from \emph{Cliff's Delta} test show that the \emph{effect sizes} are negligible (i.e., \emph{d \textless 0.1}). These findings suggest that readability changes through edits, but its overall impact remains minimal.

\begin{findingbox}
\leftskip 10pt \rightskip 5pt \textbf{Summary of RQ\textsubscript{6}}:
Text and code readability often decline through edits, though with minimal overall impact. Text readability slightly decreases (mean: 72.0 → 69.1), and code readability is poor in both versions, with readability scores of zero (initial: 22.3\% → latest: 26.4\%), and 49.7\% of edits decrease code readability. 
\end{findingbox}

% %-------------------------------------------------------------
% \section{Findings od AI-Driven Code Evaluation and Quality Judgment (RQ7)}
% \label{sec:findings-by-llm}
% %-------------------------------------------------------------

\section{Key Findings and Suggestions}
\label{sec:key-findings}

Our study uncovers critical insights into how edits impact SO answers. To enhance collaborative editing, we provide practical suggestions that can help editors, moderators, and developers ensure high-quality contributions.

\smallskip
\noindent $\textbf{--}$ \textbf{Ensure Edits Align with the Question’s Context.}
While 53.3\% of edits improve semantic relevance, 38.1\% misalign answers by adding unnecessary details or shifting focus. Editors should prioritize clarity and correctness without altering the original intent of the answer. 

\smallskip
\noindent $\textbf{--}$ \textbf{Fix the Code Without Introducing Breaking Changes.}
Edits improve 9\% of non-executable code, yet 14.7\% of previously functional code breaks after modification. This highlights the risk of well-intended but harmful changes. Users should carefully edit answers, especially when modifying logic, syntax, or dependencies.

\smallskip
\noindent $\textbf{--}$ \textbf{Avoid Unnecessary Complexity in Edits.}
While edits should simplify code, 32.3\% of edits increase complexity, making answers harder to maintain. The maintainability index declines from 86.6 to 84.2, which suggests that some edits may introduce unnecessary refactoring. Users should keep edits minimal and purposeful rather than over-engineering.

\smallskip
\noindent $\textbf{--}$ \textbf{Be Cautious About Security Risks.}
Instead of reducing vulnerabilities, 20.5\% of edits introduce more security issues. Editors should check for secure coding practices, particularly in error handling and input validation, to avoid adding vulnerabilities.

\smallskip
\noindent $\textbf{--}$ \textbf{Readability and Performance Trade-offs Need Attention.}
While some edits improve formatting, 49.7\% reduce readability, making answers harder to understand. Similarly, 51.0\% of edits optimize execution time, yet performance sometimes worsens due to additional function calls. Users should focus on clear, structured edits that improve comprehension and efficiency without adding unnecessary complexity.

\begin{table*}[!htb]
\centering
    \caption{Summary of Research Gaps and Our Contributions}
    \label{table:research-gap-summary}
    \resizebox{6.5in}{!}{%
    \begin{tabular}{p{6.5cm}|p{5.4cm}}
    \toprule
    \textbf{Existing Research Gaps} & \textbf{How Our Study Fills the Gap} \\ \midrule

    Studies focus on whether edits are accepted or rejected but do not analyze whether accepted edits enhance answer quality \cite{Wang-SOEdit-TSE2018, mondal2023automatic}. & We systematically assess how edits affect five key quality aspects (e.g., readability, usability, vulnerabilities). \\ \midrule

    Research shows that edits can reduce security risks slightly in SO C/C++ code snippets, but most of them often remain unaddressed (Zhang et al. \cite{zhang2021study}). & We extend this analysis by evaluating several other crucial aspects (e.g., bugs, code smells), including security issues in Python code edits, to identify whether edits mitigate or introduce these vulnerabilities. \\ \midrule

    Prior studies find that most edits are small and focus on surface-level changes (e.g., grammar, formatting) \cite{Baltes-SOTorrentEvolution-MSR2018}, but their impact on code quality remains unclear. & We analyze whether such edits improve or degrade code usability, complexity, and optimization, addressing this key gap. \\ \midrule

    Research has examined developer engagement with edited posts (e.g., upvotes, reputation gain) \cite{li2015good}, but does not measure whether edits improve code maintainability and efficiency. & We provide an empirical evaluation of maintainability metrics (e.g., cyclomatic complexity, Halstead complexity) to assess whether edits enhance long-term code quality. \\ \bottomrule

    \end{tabular}
    }
\end{table*}

%____________________________
\section{Threat to Validity} 
\label{sec:threatToValidity}
%____________________________

Threats to \emph{external validity} relate to the generalizability of a technique. We analyzed 94,994 answers related to the Python programming language to see whether edits enhance the quality of answers (text+code snippets) (e.g., code complexity, vulnerability). Our results may not be generalized to other programming languages. However, our findings align with those of Zhang et al. \cite{zhang2021study}, who investigated whether edits reduce Common Weakness Enumeration (CWE) vulnerabilities in C/C++ code snippets. Thus, our findings might be generalizable to other programming languages. However, we cannot guarantee the same findings for all aspects and caution readers not to over-generalize our results.

Threats to \emph{internal validity} relate to experimental errors
and biases \citep{tian2014automated}. We extract code snippets from answers using specialized HTML tags like $<$code$>$ within $<$pre$>$. However, answer submitters may include non-code content (e.g., stack traces) within these tags, which could affect our analysis of parsability, executability, and code readability. To assess the potential impact, we manually reviewed a random sample of 200 code snippets and found only three instances of non-code content. Given this minimal occurrence (1.5\%), it does not significantly affect our overall findings.

Threats to \emph{construct validity} relate to the suitability of evaluation metrics.
We use the \emph{Mann-Whitney-Wilcoxon} test, which is a widely used non-parametric test for evaluating the difference between two sample sets. However, the significance level might suffer due to the limited size of the samples. We thus consider the effect size along with the \emph{p-value}.

In addition, behavioral similarity between the initial and latest code versions could potentially affect the findings related to RQ3 (code complexity) and RQ5 (code optimization). We did not explicitly evaluate behavioral similarity because it is challenging to measure automatically and impractical to inspect 94,994 samples manually. Since these answers addressed the same questions, the behavior/intent of the two code versions remains consistent. The median edit distance of 127 characters suggests that the changes were relatively minor. To further support this, we manually examined 50 random samples and found that functional behavior and intent remained unchanged, despite code edits. Nevertheless, we acknowledge this limitation and its potential impact on our findings of code complexity and optimization.

%_______________________
\section{Related Work} 
\label{sec:relatedwork}
%_______________________

Editing is crucial in refining content on collaborative Q\&A platforms such as SO, where users modify posts to enhance clarity, correctness, and presentation. Prior research suggests that edits can improve post visibility and engagement. For instance, Li et al. \cite{li2015good} found that well-edited SO answers receive up to 119\% more upvotes, reflecting increased user appreciation. Similarly, Wang et al. \cite{Wang-SOEdit-TSE2018} reported that users tend to edit more frequently when approaching a reputation milestone, suggesting an incentive-driven motivation for editing.
Despite these findings, research remains inconclusive on whether edits truly enhance the objective quality of answers. Many edits are superficial, focusing on minor grammatical fixes or formatting rather than substantial improvements in code correctness, readability, and security \cite{Baltes-SOTorrentEvolution-MSR2018, Chen-DeepLearningCollaborativeEdit-CSCW2017}. A study by Mondal et al. \cite{mondal2023automatic} found that some edits introduce gratitude expressions and unnecessary modifications that lead to rejection. However, even accepted edits may not necessarily enhance quality. Zhang et al. \cite{zhang2021study} investigated security vulnerabilities in C/C++ snippets on SO and found that while some edits reduce weaknesses, most security issues persist. This suggests that edits may not adequately address deeper quality concerns, such as maintainability and security.

Moreover, code complexity and usability remain underexplored aspects of SO edits. Prior research has shown that high-quality code should be parsable, executable, and maintainable \cite{Calefato-HowToAskForTechnicalHelp-IST2018}. However, edits do not always simplify code; they can inadvertently increase complexity by introducing redundant logic or inconsistent modifications \cite{Wang-SOEdit-TSE2018}. While some research has examined the effectiveness of edits based on user engagement (e.g., upvotes and reputation gain), there is little empirical evidence on whether edits systematically enhance critical quality attributes such as readability, efficiency, and security.

Our study aims to bridge this gap (Table \ref{table:research-gap-summary}) by conducting a large-scale empirical analysis of SO edits, assessing their impact on five key aspects of objective quality: (1) code usability, (2) code complexity, (3) code vulnerabilities, (4) semantic relevance, and (5) readability. Unlike previous studies that focus on edit acceptance rates or user engagement metrics, we provide a quantitative evaluation of how edits impact these fundamental dimensions of answer quality.

% \textbf{Added the following paragraph}
% Through a qualitative study, Wang et al. \cite{wang2018users} identified 12 reasons (e.g., undesired code/text formatting) for rolling back a revision. Mondal et al. \cite{mondal2023automatic} further expand on this by identifying seven additional rollback reasons (e.g., adding gratitude). 
% Although these studies provide insights into the reasons for rejections, they do not assess whether accepted edits enhance critical quality aspects of answers, such as readability, code complexity, and code vulnerability--vital for efficient, reliable, and maintainable software development. Zhang et al. \cite{zhang2021study} attempt to see whether edits reduce Common Weakness Enumeration (CWE) vulnerabilities in C/C++ code snippets. Their findings indicate that while edits slightly reduce these code weaknesses, most issues remain unaddressed. Such observation demands a study that comprehensively investigates all the major aspects of objective code quality can expose further insights to users who share code snippets on SO and participate in editing so that they are aware of the potential quality issues.

%____________________________________
\section{Conclusion and Future Work}
\label{sec:conclusion}
%____________________________________

Technical Q\&A sites like SO allow collaborative editing to improve answer quality, but whether edits consistently enhance key quality aspects remains unexplored. While prior research has examined why edits get rejected, little work has assessed whether accepted edits meaningfully improve answers. One study explored C/C++ security vulnerabilities, but broader aspects like readability, code complexity, and performance remain underexplored. In this study, we analyze 94,994 edited Python answers to investigate whether edits improve six crucial quality aspects: (1) semantic relevance, (2) code usability, (3) code complexity, (4) security vulnerabilities, (5) code optimization, and (6) readability. 
Our findings show both positive and negative effects. 53.3\% of edits improve semantic relevance, but 38.1\% misalign answers with their questions. 9\% of previously broken code becomes executable, yet 14.7\% of working code breaks after edits. 32.3\% of edits increase complexity, making code harder to maintain. 20.5\% of edits introduce security vulnerabilities instead of fixing them. Readability declines in 49.7\% of cases, making answers more difficult to understand. These inconsistencies suggest that SO's editing system lacks effective quality control, and many edits fail to enhance usability, security, and maintainability. 
This study's findings can help improve SO's editing system by identifying weaknesses in how edits affect answer quality. They can inform forum designers to refine editing guidelines and moderation policies, ensuring that edits enhance usability, security, and readability.

In the future, we plan to introduce an automated tool that assesses answer revisions, detecting code complexity, code vulnerabilities, and readability in real time.

\section*{Acknowledgment}
This research is supported in part by the Natural Sciences and Engineering Research Council of Canada (NSERC) Discovery Grants program, the Canada Foundation for Innovation's John R. Evans Leaders Fund (CFI-JELF), and by the industry-stream NSERC CREATE in Software Analytics Research (SOAR).

\balance

% \begin{acks}
% This research is supported by the Natural Sciences and Engineering Research Council of Canada (NSERC), and by a Canada First Research Excellence Fund (CFREF) grant coordinated by the Global Institute for Food Security (GIFS).
% \end{acks}

% \balance
% \bibliographystyle{plainnat}
\bibliographystyle{unsrtnat}
\bibliography{reference}

\end{document}